\documentclass[12pt]{article}

\RequirePackage[OT1]{fontenc}
\RequirePackage{amsthm,amsmath}
\RequirePackage{natbib}
\RequirePackage[colorlinks,citecolor=blue,urlcolor=blue]{hyperref}
\usepackage{amssymb, amsfonts}
\usepackage{color, hyperref}
\usepackage[pdftex]{graphicx}

\title{\textbf{Statistical models for manifold data \\ with applications to the human face}}

\author{Liberty Vittert \\
School of Mathematics and Statistics \\
The University of Glasgow, Glasgow G12 8QW, UK \\
\href{mailto:liberty.vittert@glasgow.ac.uk}{liberty.vittert@glasgow.ac.uk}
\and
Adrian W. Bowman \\
School of Mathematics and Statistics \\
The University of Glasgow, Glasgow G12 8QW, UK \\
\href{mailto:adrian.bowman@glasgow.ac.uk}{adrian.bowman@glasgow.ac.uk}
\and
Stanislav Katina \\
Institute of Mathematics and Statistics \\
Masaryk University, Brno, Czech Republic \\
\href{mailto:katina@math.muni.cz}{katina@math.muni.cz}
}

\begin{document}

\maketitle

\begin{abstract}
One of the data structures generated by medical imaging techno\-logy is high resolution point clouds representing anatomical surfaces.  Stereo-photogrammetry and laser-scanning are two widely available sources of this kind of data.  Raw images are in the form of triangulated surfaces and the first step is to create a standardised representation of surface shape which provides a meaningful correspondence across different images, to provide a basis for statistical analysis.  Point locations with anatomical definitions, referred to as landmarks, have been the traditional approach, with analysis and interpretation which is widely used and well understood.  Landmarks can also be taken as the starting point for more general surface representations, often using templates which are warped on to an observed surface by matching landmark positions and subsequent local adjustment of the surface.  The aim of the present paper is to use the intermediate structures of ridge and valley curves to capture the principal features of the manifold (surface) of interest.  Landmarks are used as anchoring points, as usual, but curvature information across the surface is used to guide the curve locations.  Once the ridges and valleys have been located, the intervening surface patches are relatively flat and can be represented in a standardised manner by appropriate surface transects, to give a complete surface model.  However, the intermediate curve representation is of considerable interest in its own right, as it captures the principal features of the surface and therefore embodies much of the shape information on the object of interest.

Methods of identifying these curves are described and evaluated.  While the methods are generic, the human face is used as an important application.  In particular, the models are used to investigate sexual dimorphism (the differences in shape between male and female faces).  The curve representation is shown to capture a major component of the relevant information.
\end{abstract}


\section{Introduction}
\label{sec:introduction}

One of the interesting types of data generated by medical imaging is in the form of anatomical surfaces.  These can arise through the thresholding of three-dimensional voxel data, for example to identify the transition between soft tissue and bone.  They can also be generated through techniques such as laser scanning or stereo-photogrammetry, where surface locations are measured directly through optical methods.  Each observational unit consists of a three-dimensional point cloud, with an associated triangulation, which provides a discrete, and to some extent noisy, representation of the target surface or manifold.

An immediate issue in the analysis of manifold data is the need to create standardised (\textit{homologous}) representations which have anatomical and geometrical correspondence across images.  A traditional approach to this has been through landmarks, which identify key point locations; see, for example, \cite{farkas-1994-book} for a discussion of this in the context of the human face.  There is, necessarily, substantial loss of information in reducing the representation to a small number of points, however well chosen, but the wide availability of statistical tools for the analysis of this type of shape information has given landmarks a central role in the study of shape.  \cite{dryden-1998-book} give an authoritative account of this topic.

At the other end of the spectrum, a variety of approaches can be taken to build full surface representations.  The majority of these take landmarks as a starting point.  A common approach employs a surface template which is deformed, or warped, so that landmark positions on the template match exactly those on the manifold of interest.  \cite{hammond20043d} used this approach to develop a \textit{dense point correspondence} model where standar\-dised non-landmark locations on the image of interest are identified as the closest points to the corresponding positions on the warped template.  This approach was developed further by \cite{zhao-2011-smc} by adding information on local surface curvature to guide further deformation of the template and increase the quality of the match with the manifold in geometrical terms.  In a different approach, applied to human faces, \cite{srivastava2009elastic} use a Riemannian framework to produce a co-ordinate system which allows both deformation and comparison using a single elastic metric.  A Darcyan curvilinear co-ordinate system, based on the geodesic distance function from a fixed reference point, allows the facial surface to be represented as an indexed collection of level curves.

The present paper tackles the problem from a different perspective by focusing attention on anatomical curves which identify ridges and valleys.  These are intermediate structures which provide a richer description of manifold data than landmarks but which, as inherently one-dimensional objects, provide simpler representations than a full surface approach.  Ridge and valley curves often capture many of the key features of a manifold, as discussed by \citet{koenderink-1990-book} and many other authors.  For example, in the context of the human face a large proportion of the information on shape is captured in the locations of the ridges, such as the nose profile, and valleys, such as where closed lips meet.  Curves may incorporate anatomical landmarks; indeed, \cite{katina-2015-janatomy} argue that the definitions of anatomical landmarks should be based on anatomical curves, through crossing points or the locations with maximum curvature.  Full surface representations can be approached by relatively straightforward in-filling of areas between curves, as the curvature of these intervening regions will be relatively low.  There are therefore strong arguments that ridge and valley curves are the key features to target in constructing representations of shape for manifolds.

In general, the estimation of the locations of ridge and valley curves on a manifold is a difficult problem, in the absence of prior information on the nature of the curves to be located.  One broad approach is to exploit the definition of ridge or valley points in terms of locally extreme curvature and then to combine candidate points into a curve.  \cite{ohtake2004ridge}, \cite{stylianou-2004-ieee} and \cite{che2011ridge} provide examples of this approach from the extensive computer vision literature on this topic.  \citet{bowman-2015-csda} used repeated scans along surface transects to locate points of discontinuity in first derivatives, or sharp change in direction, using statistical methods.  Again, ridge and valley curves were created by combining candidate points, although the principal curve \cite{hastie-1989-jasa} methods employed do not guarantee a curve which lies on the manifold.

The problem of ridge and valley curve estimation is tackled in the present paper by a new approach which uses basic information on surface curvature to identify regions of interest, transforms these locally to a two-dimensional space, and exploits the expected smoothness of the curves to estimate these directly in this new domain.  Although the underlying methods have considerable generality, the human face is used as a significant application.  This is a context where anatomical landmarks are widely available through trained manual identification or through more automatic methods, as described for example by \cite{sukno-2014-cybernetics}.  The problem of estimating ridge and valley curves is therefore tackled in the setting where landmarks are available to provide end points of each curve of interest.

Elementary ideas of differential geometry are reviewed in Section~\ref{sec:diffgeom} to provide some basic tools for characterising manifolds.  Methods for the estimation of ridge and valley curves are described in detail in Section~\ref{sec:estimation}, using the human face as an important example.  The construction of a full surface model is then addressed in Section~\ref{sec:mesh}, where the concept of `sliding' to ensure geometric matching with a template is discussed.  The properties of the method are evaluated computationally, and in a simulation study, in Section~\ref{sec:evaluation}.  The resulting models are used to explore human sexual dimorphism in Section~\ref{sec:application}.  Some final discussion is given in Section~\ref{sec:discussion}.

\section{Tools from differential geometry}
\label{sec:diffgeom}

The local shape at a three-dimensional location $m=(x,y,z)$ on a differentiable manifold $\mathcal{M}$ can be characterised through the quadratic surface
\begin{equation}
   z = \frac{1}{2} \left(\kappa_1 x^2 + \kappa_2 y^2\right),
\label{eq:quadratic}
\end{equation}
where $z$ lies in the normal direction to the surface and the orthogonal axes $x$ and $y$ lie on the tangent plane, associated with the directions of maximum ($\kappa_1$) and minimum ($\kappa_2$) curvature.  This is clearly described by \citet{koenderink-1990-book} and many others, along with a wide variety of other key tools for studying surface shape.  The \textit{principal curvatures}, $\kappa_1$ and $\kappa_2$, and their associated directions, provide the essential information for characterising curvature across the manifold.

An observation of a manifold consists of a three-dimensional point cloud plus an associated triangulation.  In order to estimate the principal curvatures at an observed point $m$, a surface normal direction can be constructed as the average of the normal vectors associated with the set of triangles which contain $m$; see \cite{koenderink-1992-imagevisioncomputing}.  By fitting a quadratic model of the form (\ref{eq:quadratic}) to a local neighbourhood of points through ordinary least squares, with arbitrary orthogonal axes in the tangent plane, the principal curvatures and directions can be estimated through the eigen-decomposition of the Weingarten matrix.  \cite{goldfeather2004novel} provide all the details.  An important choice is the size of the neighbourhood to which the quadratic surface is fitted.  In general, this needs to be adapted to the characteristics of the manifolds being studied.  Working with human faces which, at a broad level, have very strong shape correspondences across people, allows the effects of different neighbourhood size to be explored.  This resulted in the recommendation of using connected triangulation points within a radius of $1\,$cm of the point of interest.

\begin{figure}[htp]
\centerline{
   \includegraphics[width= 1.0\textwidth]{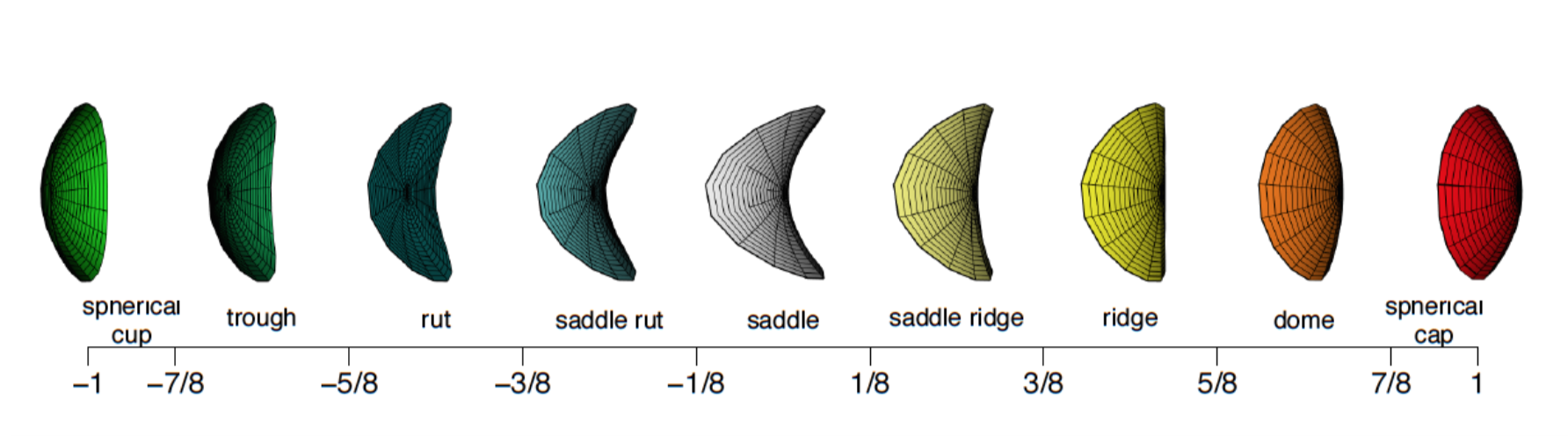}}
\centerline{
   \includegraphics[width= 0.4\textwidth]{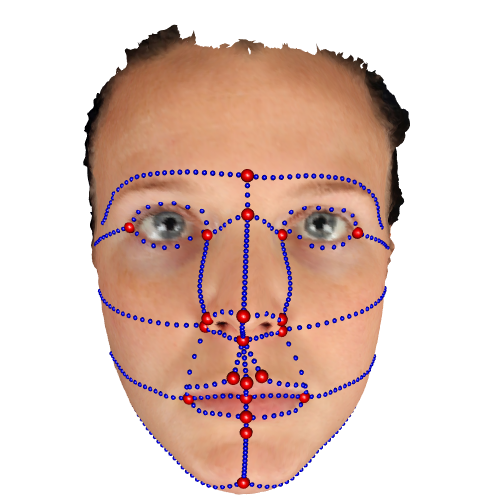}
   \includegraphics[width= 0.4\textwidth]{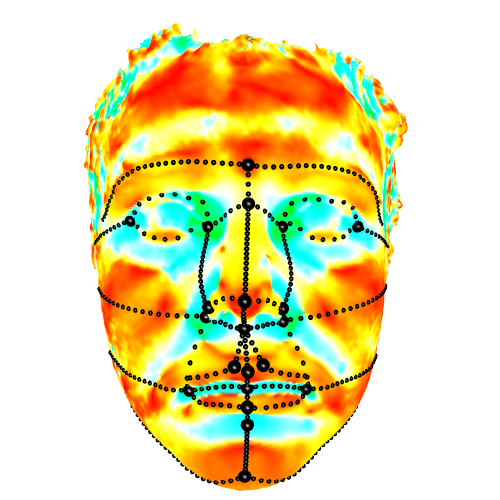}}
\caption{The upper plot illustrates the local surfaces associated with the shape index on the scale from $-1$ to $1$, with colour coding to identify each shape category.  The lower images show manually placed anatomical landmarks ($23$ points), and anatomical curves ($467$ points), on facial surfaces coloured by natural texture (left) and shape index (right).}
\label{fig:shapeindex}
\end{figure}

There are many useful ways in which the principal curvatures can be condensed into a single number, to express particular curvature properties.  For example, \textit{Gaussian curvature}, defined as $\kappa_1 \kappa_2$, is a very commonly used measure where positive values correspond to peaks, wells, ridges or valleys, while negative values correspond to saddle points.  A particularly useful summary in the present setting is the \textit{shape index} \citep{koenderink-1992-imagevisioncomputing} which gives a helpful characterisation of the type of curvature at any point.  This is defined as
\begin{equation}
   S(m) = \frac{1}{2} - \frac{1}{\pi} \tan^{-1} \left( \frac{\kappa_1(m) + \kappa_2(m)}{\kappa_1(m)-\kappa_2(m)} \right),
\end{equation}
where the notation emphasises that the principal curvatures vary with the point of interest $m$.  Values of $S$ close to $-1$ indicate a `spherical cup' where both principal curvatures are positive.  As $S$ increases the corresponding surface shape bends smoothly through `trough' and `rut', reaching `saddle' shapes around $0$.  The process is reversed as $S$ increases and the associated surfaces move through `ridge' and `dome' to arrive at a 'spherical cap' when $S$ reaches $1$.   This helpful typology of surface shape is illustrated in the upper part of Figure~\ref{fig:shapeindex}, which is modelled on a similar figure in \citet{koenderink-1992-imagevisioncomputing} and uses the same verbal descriptors and colour coding.  Notice that the shape index $S$ is a function of the ratio of $\kappa_1$ and $\kappa_2$, so that it describes the type of local curvature, but not its strength. 
 
The lower images of Figure~\ref{fig:shapeindex} show a human face with the right hand image coloured by shape index.  This highlights key features of the face, such as the ridge of the nose and the valleys and wells around the eyes.  The positions of manually placed landmarks and curves align well with key curvature features.  This provides encouragement for the use of the shape index, and principal curvatures more generally, in the estimation of anatomical curves.  This topic is developed in the following section.

\section{Estimation of ridge and valley curves}
\label{sec:estimation}

Working in a space defined by a manifold requires careful adaptation of the standard procedures used in linear spaces.  \citet{patrangenaru-2015-book} discuss this in detail.  A simple example is the calculation of an empirical mean over a manifold $\mathcal{M}$ represented by a set of three-dimensional points $\{m_i = (x_i, y_i, z_i); i = 1, \ldots, n\}$ and the associated triangulation.  An empirical \textit{Fr\'echet mean} is defined implicitly as $\mbox{arg} \min\limits_{\forall \,m} \sum_{i=1}^n d(m, m_i)^2$, where $m$ lies in the manifold and the function $d$ measures distance within the manifold (not in Euclidean space).  So, in seeking to estimate curves, the methods adopted should also seek to ensure that these estimates lie within the given manifold.

To establish notation, it will be useful to refer to a curved path over a manifold as $p(s) = \{x(s), y(s), z(s)\}$, where the functions $x, y, z$ describe the movement of the three-dimensional co-ordinates as functions of an arc length argument $s$.  In terms of the observed manifold, this is expressed in the set of locations $\{p_j = (x_{pj}, y_{pj}, z_{pj}); j = 1, \ldots, n_p\}$ where the curve intersects the triangulation.  It is also helpful to provide a characterisation of a ridge (or valley) curve.  \cite{kent-1996-bmvc} provides a clear approach to this through the concept of `principal curves' on a surface, meaning curves where, at every point, the tangential directions coincide with a direction of principal curvature of the surface.  The points along these principal curves at which the curvature is locally extreme are called ridge points.  The set of all such points forms ridge curves.
    
\begin{figure}[htp]
\centerline{
   \includegraphics[width = 0.3\textwidth]{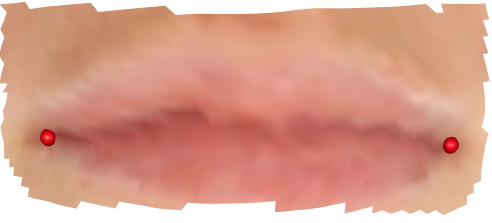}
   \includegraphics[width = 0.3\textwidth]{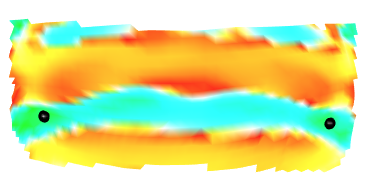}
   \includegraphics[width = 0.3\textwidth]{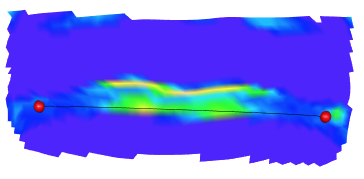}}
\centerline{
  \parbox{0.35\textwidth}{
  \vspace{-24mm}
   \includegraphics[width=0.35\textwidth]{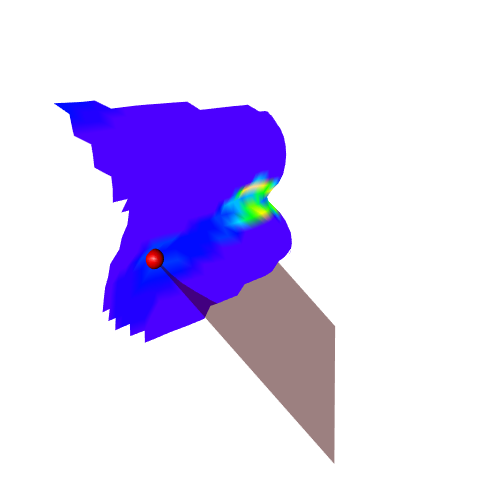}}
   \includegraphics[width=0.31\textwidth]{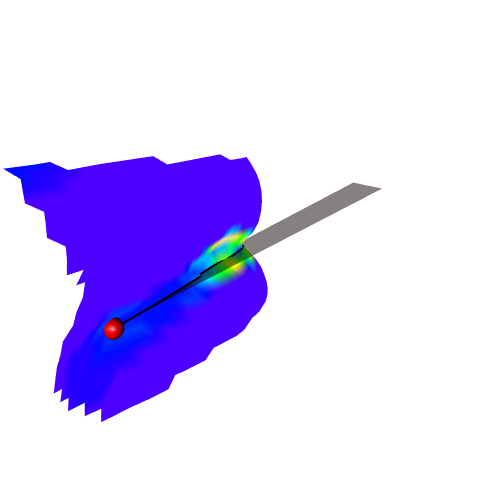}
   \includegraphics[width=0.31\textwidth]{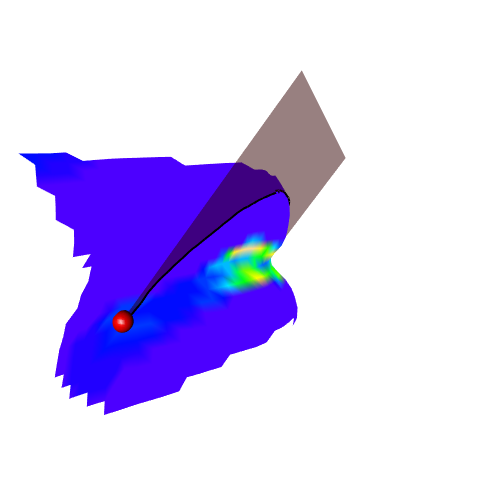}}
\caption{The upper plots show a three-dimensional manifold of lips coloured by texture (left), shape index (middle) and maximum curvature values (right).  The lower plots show three plane cuts to determine the linear path which maximises the length-standardised integral of maximum curvature.}
\label{fig:plane-path}
\end{figure}

In order to illustrate the issues involved, consider the case of estimating the valley curve where closed lips meet in a human face, illustrated in Figure~\ref{fig:plane-path}, with two standard landmarks locating the corners of the mouth.

\subsection{A linear reference path}
\label{sec:linear-path}

There is no natural co-ordinate system on the manifold, so the creation of a reference path between the two landmarks ($l_1$, $l_2$) is a helpful first step.  The strategy, defined below in detail, is to identify a local region of interest, use shape index to identify the location of relevant surface shapes, and then identify the planar surface cut whose path maximises the curvature across the transect.
\begin{enumerate}
\item   \textit{Localisation by distance.}
The presence of two reference landmarks allows attention to be focussed on a local region of the manifold.  For human faces, a strong degree of smoothness in ridge and valley curves is to be expected, quantified in the requirement that all points on the curve  between $l_1$ and $l_2$ should lie within the cylinder with axis $(l_2 - l_1)$ and radius $r$.  Empirical investigations with human faces led to the choice $r = ||l_2 - l_1||/2$ as an effective method of identifying the local region of interest.  The top left panel of Figure~\ref{fig:plane-path} gives an example of the localised region for the lips.
\item   \textit{Localisation by surface shape.}
The shape index $S$ over the manifold, illustrated in the top middle panel of Figure~\ref{fig:plane-path}, can now be used to identify those points within the local manifold which indicate valley behaviour.  In order to be inclusive at this stage, all points with negative shape index, corresponding to blue-green colour in the standardised scale of \citet{koenderink-1992-imagevisioncomputing}, are included.  
\item   \textit{Localisation by curvature.}
As observed above, shape index describes the type of surface shape but not its strength.  For a valley, strength can be quantified by $\nu = \max(\kappa_1, \kappa_2)$ which we know will be positive.  Non-valley points, with positive shape index, can be assigned the value $0$.  The top right hand panel of Figure~\ref{fig:plane-path} illustrates the result using topographic colour coding.
\end{enumerate}
A reference path can now be identified by considering the set of planes containing $l_1$ and $l_2$, indexed by an angle of orientation $\gamma$.  The lower panels of Figure~\ref{fig:plane-path} illustrate this.  For each $\gamma$, the intersection of the plane with the manifold describes a path $p_{\gamma}$ from $l_1$ to $l_2$.  In view of the characterisation of ridge points as positions of locally extreme curvature, the integral along a ridge curve inherits this locally extreme property.  The integral along any other nearby curve of the same length must be smaller.  This provides a mechanism for identifying a reference path through optimising integrated curvature.  The standardised integral of the maximum curvature along the path $p_{\gamma}$ is
$$
   \left\{ \int_{p_{\gamma}} \nu(s) ds \right\} / \left\{ \int_{p_{\gamma}}1 \, ds \right\} \approx
                      \left\{ \sum_{j=2}^{n_p} w_j \nu(p_j) \right\} / \left\{ \sum_{j=2}^{n_p} w_j \right\} ,
$$
where the right hand expression shows the discrete approximation based on the path intersection points $\{p_j; j = 1, \ldots, n_p\}$ on the observed manifold and the weights $w_j = ||p_j - p_{j-1}||$ measure the distances between successive intersection points.  The curve associated with the $\gamma$ which maximises this expression then `mops up' as much curvature as possible.  The standardisation by curve length $\int_{p_{\gamma}}1\, ds$ is required to penalise curves which `pick up' large amounts of curvature by travelling long distances.

The resulting curve is `linear' in the sense that it lies in a two-dimensional plane embedded in three-dimensional space, but it describes a curved path over the manifold.  The black points in the left hand panel of Figure~\ref{fig:3d2d} illustrate the resulting path for the lip valley example.

There are occasions when it is useful to identify a curve across a region which has little strong curvature, such as the cheek from ear to nose landmarks.  A linear reference path can easily be constructed simply by minimising over$\int_{p_{\gamma}} 1 \, ds \approx \sum_{j=2}^{n_p} w_j$ to locate the planar cut with minimum length.

\subsection{A more flexible estimate}

\begin{figure}
\parbox[t]{0.3\textwidth}{\hrule height 0pt width 0pt \vspace{10mm}
   \includegraphics[width=0.3\textwidth]{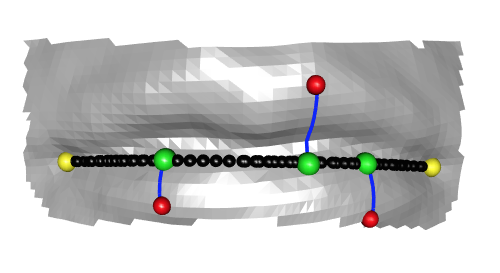}}
\parbox[t]{0.3\textwidth}{\hrule height 0pt width 0pt \vspace{10mm}
   \includegraphics[width=0.3\textwidth]{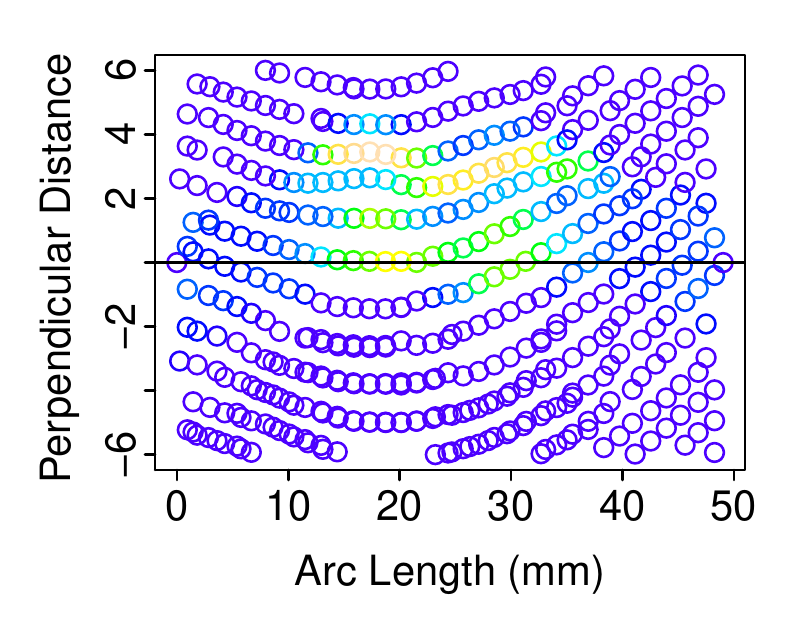}}
\parbox[t]{0.3\textwidth}{\hrule height 0pt width 0pt  \vspace{10mm}
   \includegraphics[width=0.3\textwidth]{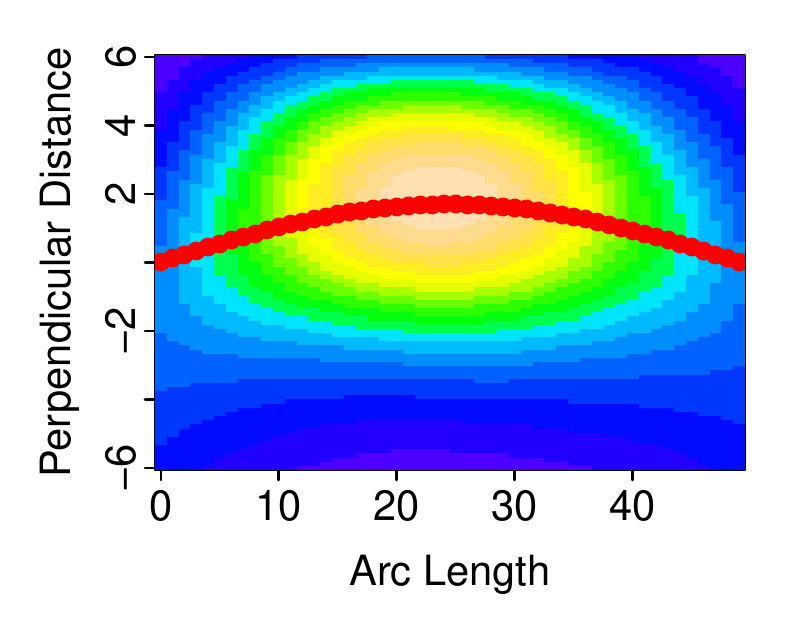}}
\caption{The left hand image shows lips with anatomical landmarks (yellow), a linear reference curve (black), some illustrative points on the manifold (red), and the corresponding closest (perpendicular) distances to the reference curve (blue).  The middle image shows the points from the manifold in the two-dimensional space defined by signed distance to the closest point on the reference curve and arc length along the reference curve to the location of the closest point.  The colour coding of the points shows the relative strengths of the maximum curvature at each location, on a topographic scale.  The right hand image shows a smooth surface representation, with the flexible estimate of the ridge curve in this two-dimensional space indicated by the red line.}
\label{fig:3d2d}
\end{figure}

The planar constraint on the reference path described above means that it can only be regarded as a first approximation to an estimate of the valley curve of interest.  However, the existence of this reference allows the construction of a more flexible estimate through the creation of a local co-ordinate system which uses the reference path as a baseline.  The intrinsically two-dimensional nature of the manifold can be represented by two co-ordinate axes: one relates to the signed distance of each point on the manifold from its closest point on the reference path ($d$), while the other describes the arc length along the reference path of these closest points ($s$).  This is illustrated in the left hand panel of Figure~\ref{fig:3d2d}.  The distance of a manifold point to the closest point on the reference path can be measured along the manifold but on this occasion the simple Euclidean distance is a very good substitute because of the localisation to a region of the manifold very close to the valley curve of interest.  The middle panel of Figure~\ref{fig:3d2d} gives an example of the manifold points represented in this new two-dimensional space.

In earlier discussion, a valley point was characterised as a location of maximal curvature along the direction of principal curvature.  In more informal language, the curvature `across' the valley is maximised at the valley points.  The two-dimensional axes created through the reference curve, as described above, then give the opportunity of identifying the valley by locating the positions of these maximum curvatures.  The vertical axis, corresponding to the distance of each mesh point from the reference curve, will not be aligned exactly with the direction of maximum principal curvature for each point.  Consider the baseline case of a valley formed from the surface $z = f(x, y) = \beta y^2$.  It is straightforward to derive the two principal curvatures at the point $(x, y)$ as $\kappa_1 = 2\beta/(1 + 4 \beta^2 y^2)^\frac{3}{2}$ and $\kappa_2 = 0$.  It is then clear that the valley points remain the locations of maximum local curvature along any direction other than the valley curve itself ($y = 0$).  More generally, the curvature patterns are affected by changes in the strength of the quadratic shape across the valley and by the movement of the position of the valley curve in the $(x, y)$ plane and in the $z$-axis.  Modest changes in these features, such as those exhibited in human faces, will not dislodge the valley points as positions of local maximum curvature.  This is explored quantitatively in the Supplementary Material.

In order to estimate the location of the valley curve at any particular arc length, it is therefore necessary to estimate where the curvature $\nu$ is maximised across the $d$-axis in the new $(s, d)$ domain.  (Recall that $s$ is arc length and $d$ is perpendicular distance.)  The middle panel of Figure~\ref{fig:3d2d} uses colour shading to display the values of the larger principal curvature ($\nu_i$), using the two-dimensional locations $(s_i, d_i)$ for each mesh point $(i=1, \ldots, n)$ where the shape index is negative.  This shows strong valley curvature in the central region with weaker curvature near the corners of the mouth.

The curvature surface can be conveniently represented in two-dimensional p-spline form \citep{eilers1996flexible} as
$$
   \nu(s, d) = \sum_i \sum_j \hat{\beta}_{ij} \phi_i(s) \phi_j(d) ,
$$
constructed from the product of two one-dimensional cubic b-spline bases $\{\phi_i; i = 1, \ldots, b\}$ and a set of basis weights $\{\hat{\beta}_{ij}; i, j = 1, \ldots, b\}$.  The presence of the `hat' symbol on each $\hat{\beta}_{ij}$ reflects the fact that these have been estimated by fixing the equivalent degrees of the estimator to be $12$, which allows a good degree of flexibility for the fitted surface.  The aim is to locate the surface ridge curve, for which a one-dimensional p-spline representation could also be used.  However, an even simpler approach is to represent the ridge curve $r$ at a grid of arc length positions $\{s_k; k = 1, \ldots, n_g\}$ as
$$
   r(s_k) = \alpha_k, \ \ \ k = 1, \ldots, n_g .
$$
Following the earlier principle that the integral of curvature along a ridge curve is locally maximal, the aim is to identify $r(s)$ to maximise
$$
   \int \nu(s, r(s)) ds .
$$
By adopting a discrete approximation, and by exploiting the assumption of smoothness in the ridge curve through a penalty function, this translates into the problem of identifying the $\mathbf{\alpha}$ (the vector of $\alpha_k$ values) which maximises the penalised discrete integral
\begin{eqnarray*}
   M & = & \frac{1}{n_g} \sum_k \nu(s_k, \alpha_k) - \lambda \mathbf{\alpha}^T P \mathbf{\alpha} \\
       & = & \frac{1}{n_g} \sum_k \sum_i \sum_j \hat{\beta}_{ij} \phi_i(s_k) \phi_j(\alpha_k) 
                       - \lambda \mathbf{\alpha}^T P \mathbf{\alpha},
\end{eqnarray*}
where $\lambda$ denotes a penalty parameter and $P = D^TD$ denotes the penalty matrix constructed from the matrix $D$ which generates second-order differences of the elements of $\alpha$ as $D \alpha$.  As the expression $M$ is not linear in $\alpha$, Newton's method provides a suitable solution.  This requires the evaluation of the derivatives
\begin{eqnarray*}
   \frac{\partial M}{\partial \alpha_k} & = & \sum_i \sum_j \hat{\beta}_{ij} \phi_i(s_k) \phi'_j(\alpha_k) -
               2 \lambda (P \alpha)_k , \\
   \frac{\partial^2 M}{\partial \alpha_k^2} & = & \sum_i \sum_j \hat{\beta}_{ij} \phi_i(s_k) \phi''_j(\alpha_k) -
               2 \lambda (P)_{k,k} , \\
   \frac{\partial^2 M}{\partial \alpha_k \partial \alpha_{l}} & = & -2 \lambda (P)_{k,l} , \ \ \ (k \neq l), 
\end{eqnarray*}
where the notation $(v)_k$ denotes the $l$th element of the vector $v$ and $(A)_{k,l}$ denotes the $(k,l)$th element of the matrix $A$.  B-splines have the very convenient property that their derivatives are scaled b-splines of lower order.  Specifically,
$$
   \phi'_{i,a}(x) = (\phi_{i,a-1}(x) -  \phi_{i+1,a-1}(x)) / b ,
$$
where the second subscript on $\phi$ denotes the order of the b-spline function and $b$ denotes the distance between the (regularly spaced) knots.  Second derivatives can then be computed by a further application of this formula.

The anchoring points correspond to $\alpha_1 = 0$ and $\alpha_{n_k} = 0$, so these two elements are fixed.  From a starting point which sets all $\alpha_k = 0$, Newton's method then employs the iterations
$$
   \mathbf{\alpha}^{(m+1)} = \alpha^{(m)} - H^{-1}(\alpha^{(m)}) f(\alpha^{(m)}) ,
$$
where $f$ denotes the vector of first derivatives and $H$ denotes the matrix of second derivatives.  The iterations converge very quickly.

After experimentation on a variety of images, the penalty parameter was set to $\lambda = 0.5$ as this achieves a good compromise between fidelity and smoothness for the relatively simple curves in human faces. The right hand panels of Figure~\ref{fig:3d2d} show the end result of this process on the illustrative example of a lip valley curve.  The resulting curve has a flexibility which is modest, but sufficient to adapt to the curvature information which suggests that the true valley curve deviates a little from the earlier linear reference path.

\subsection{Interpolation back to three-dimensional space}
\label{subsec:interp}

The final step is to transfer the estimate of the valley curve from the $(s, d)$ domain back to three-dimensional space.  The existence of a triangulation on the observed manifold means that barycentric interpolation can provide a simple solution.  The observed manifold usually has an associated triangulation and where this does not exist Delaunay triangulation \citep{deberg-2000-book} can be used.  Any interior point $m$ in the $(s, d)$ domain can be expressed as a linear combination of the vertices $v_1, v_2, v_3$ of its enclosing triangle, so that $m = \sum_{i=1}^3 \delta_i v_i$.  As each vertex has an associated location in three-dimensional space, the $x$-coordinate of $m$ can be interpolated as $x_m = \sum_{i=1}^3 \delta_i x_i$, where $x_1, x_2, x_3$ are the $x$-coordinates of $v_1, v_2, v_3$ in three-dimensional space.  This can then be repeated for the $y$ and $z$-coordinates.  \citet{meyer2002generalized} provide the details.  The locations of any points on the valley curve identified in the $(s, d)$ domain can therefore easily be interpolated back to three-dimensional space.

\begin{figure}
\parbox[t]{0.6\textwidth}{\hrule height 0pt width 0pt
\includegraphics[width=0.6\textwidth]{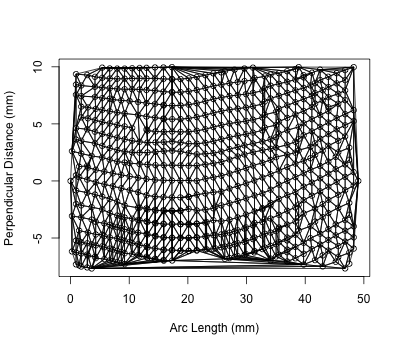}
}
\parbox[t]{0.3\textwidth}{\hrule height 0pt width 0pt
\includegraphics[width=0.3\textwidth]{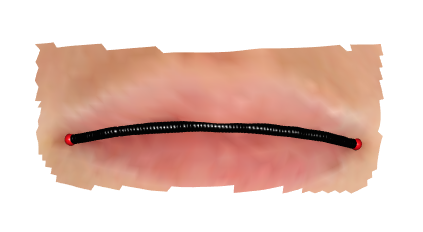}
\includegraphics[width=0.3\textwidth]{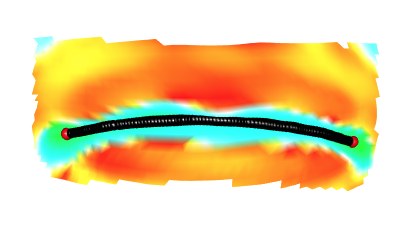}
\includegraphics[width=0.3\textwidth]{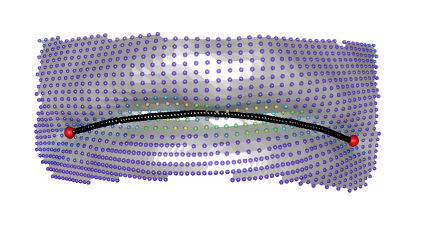}
}
\caption{The left hand panel shows a Delaunay triangulation which forms the basis of barycentric interpolation.  The right hand images show the final estimate of the lip valley curve, on surfaces coloured by natural texture, shape index and maximum curvature.}
\label{figure:interp}
\end{figure}

Figure~\ref{figure:interp} illustrates this process.  The images on the right hand side display the resulting estimate of the lip valley curve on surfaces which are coloured by shape index and maximum curvature, as well as by natural texture, in order to highlight the information which has been used in the construction of the estimate.  This process can be repeated across all the curves of the face displayed in Figure~\ref{fig:shapeindex}.  In fact, this Figure shows the curves estimated by the methods described in this section.

\section{Construction of a facial surface model}
\label{sec:mesh}

A set of anatomical curves across the face creates a structure which already captures a considerable amount of the information on shape.  A full representation of the facial surface can then be produced by characterising the patches which have these curves as boundaries.  As the construction of the curves targets locations of high curvature, these patches will necessarily have lower curvature but they may still express important biological signals.  Discrete representation of the anatomical curves, using suitably spaced points along each, then allows a description through linear transects between these boundary points, using plane cuts minimised over length as described in Section~\ref{sec:linear-path}.  

However, care must be taken in the construction of a discrete representation.  This is an issue which often occurs in functional data analysis, for example where growth curves exhibit peaks at different times across individuals.  Naive averaging across curves at each time point blurs the signals displayed in each individual curve.  \citet{kneip-2008-jasa} show how `structural averaging' can be achieved by employing a transformation of the timescale for each curve to align on key features.  In the context of shape, corresponding issues arise and \citet{bookstein-1997-mia} describes a related idea of `sliding landmarks' where the positions of points are successively adjusted along the curve to minimise bending energy with respect to a template curve.  This is a widely used technique in geometric morphometrics and \citet{rohr-2001-book} discusses the connection with principal curvatures.  

These ideas are now adapted to the present context.  In contrast to earlier work, the adjusted discrete approximation always remains on the manifold rather than using tangent approximations and back-projection.  A template is used to adjust curves successively, to create a full facial surface model.  Adjustments are then iterated across the sample of images, using the mean shape as a reference, to ensure that the locations of the adjusted points have the same interpretation across the images with respect to the geometrical features of the shapes.  This is what is meant by geometric homology.

\begin{figure}[htp]
\centerline{
\includegraphics[width= 0.5\textwidth]{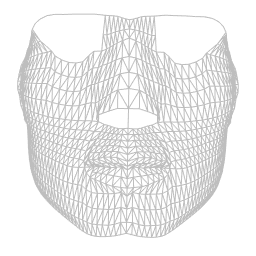}
\includegraphics[width= 0.5\textwidth]{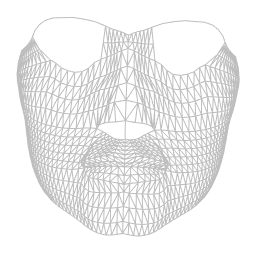}}
\caption{Created template facial triangulation for males (left) and females (right) scaled for visualisation purposes. }
\label{figure:template}
\end{figure}

The starting point is a template shape consisting of a completely symmetric, dense facial surface.  Figure~\ref{figure:template} shows examples, separately for males and females to respect potential differences in shape between the sexes for visualisation purposes.  A template can be created by adopting a particular image, using the methods described in Section~\ref{sec:estimation} to locate anatomical and intermediate curves, and hence a discrete representation  through points which are equally spaced along the arc length of each curve.  Perfect symmetry is then imposed by averaging the new image representation and its Procrustes aligned relabelled reflection.  This provides a reference to guide the adjustment of the discrete representation of other images.

The ideas behind adjustment are described first with a single curve denoted by $\mathbf{p}_I(s) = (x_I(s), y_I(s), z_I(s))$ and the corresponding template curve by $\mathbf{p}_T(s) = (x_T(s), y_T(s), z_T(s))$.  If the mapping from the template curve to the target curve is denoted by $\mathbf{p}_I(s) = \mathbf{f}(\mathbf{p}_T(s))$ then the \textit{bending energy} of the function $\mathbf{f} = (f_x, f_y, f_z)$ is
$$
   J\left( \mathbf{f}\right) = \sum_{k=x,y,z} \int \int \int_{\mathbb{R}^3} \left[ \sum_{a,b = x,y,z} \left( \frac{\partial^2f_k}{\partial a \partial b} \right) ^2\right] dx dy dz.
$$
\cite{bookstein-1989-ieeepami} discuss this in detail in the two-dimensional case, while \cite{gunz-2005-inbook} discuss the three-dimensional case.  As it is based on second derivatives, bending energy is unaffected by translation or orientation.  $J\left( \mathbf{f}\right)$ will be large if the function$\ \mathbf{f}$ exhibits high local curvature, because this will result in large second derivatives.  Any transformation $\nu(s)$ on the arc length parameter scale $s$ for the image, so that $\mathbf{p}_I(\nu(s)) = \mathbf{f}(\mathbf{p}_T(s))$, will affect the bending energy of $\mathbf{f}$.  The aim is therefore to identify the function $\nu(s)$ which minimises this bending energy, with the aim of making the curvature of the two curves as close as possible on this new scale.  The image and the template are then geometrically homologous.

The starting point is a set of interpolated values which are equally spaced along the image curve $\left\{\mathbf{p}_{Ij} = \left(x_{Ij}, y_{Ij}, z_{Ij}\right); j = 1, \ldots, n_I\right\}$ and the corresponding set of equally spaced points $\left\{\mathbf{p}_{Tj} = \left(x_{Tj}, y_{Tj}, z_{Tj}\right); j = 1, \ldots, n_I\right\}$ on the template curve.  The use of a thin-plate spline ({\sc tps}) to encapsulate the mapping between these sets of points is particularly convenient because this is designed specifically to minimise the bending energy $J\left( \mathbf{f}\right)$ associated with interpolation of the observed data. The aim is then to minimise this bending energy $J\left( \mathbf{f}\right)$ at a further level by adjusting the locations $\{\mathbf{p}_{Ij}\}$.  In an iterative algorithm, each location $\mathbf{p}_{Ij}$ on the image curve (other than the anchoring end-points) is allowed to move along the curve, with its new location $\mathbf{p}^*_{Ij}$ defined by
$$
      \mathbf{p}^*_{Ij} = \arg \min_{\mathbf{p}^*_{Ij}} J\left( \mathbf{f^*} \right) ,
$$
where $\mathbf{f}^*$ denotes the {\sc tps} function based on the adjusted observations.  This process is repeated successively across all the points on the image curve until convergence.  After all anatomical curves have been adjusted, intermediate curves are then computed as plane cuts using the new anatomical curve locations as end-points.  These intermediate curves are then also adjusted by sliding with respect to the template.

\begin{figure}
\centerline{
   \includegraphics[width= 0.32\textwidth]{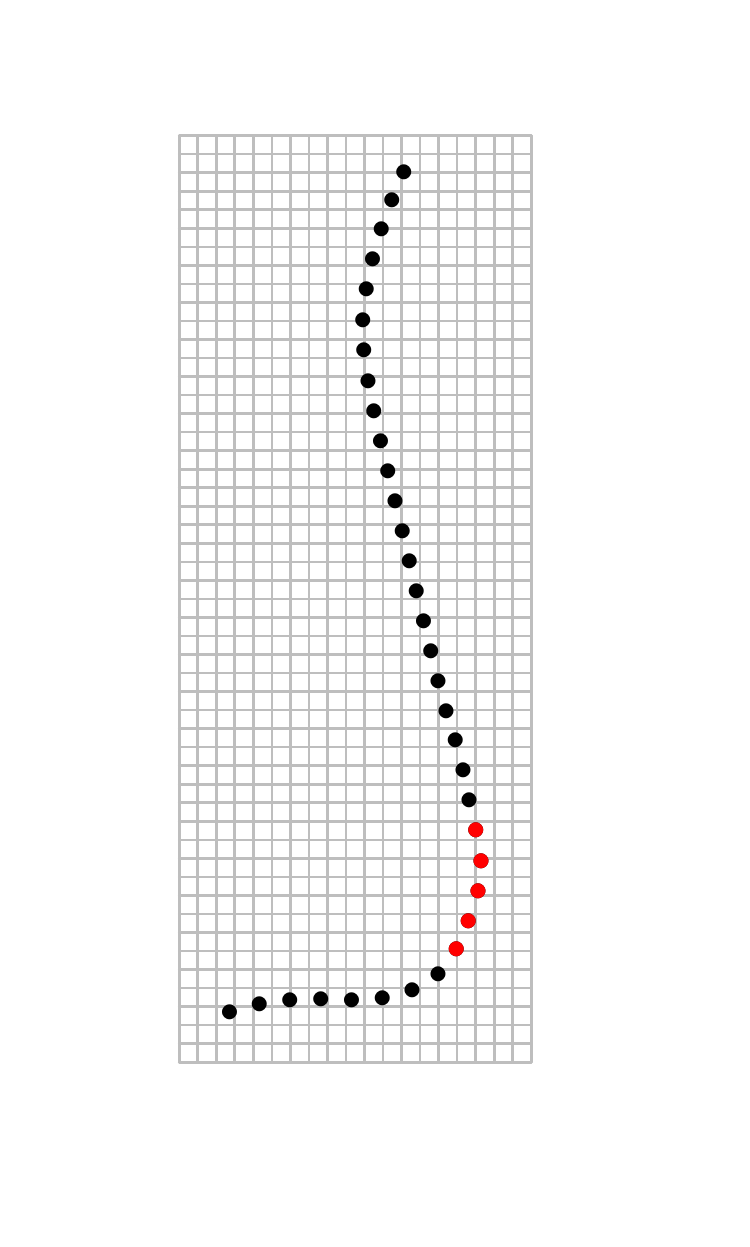}
   \includegraphics[width= 0.66\textwidth]{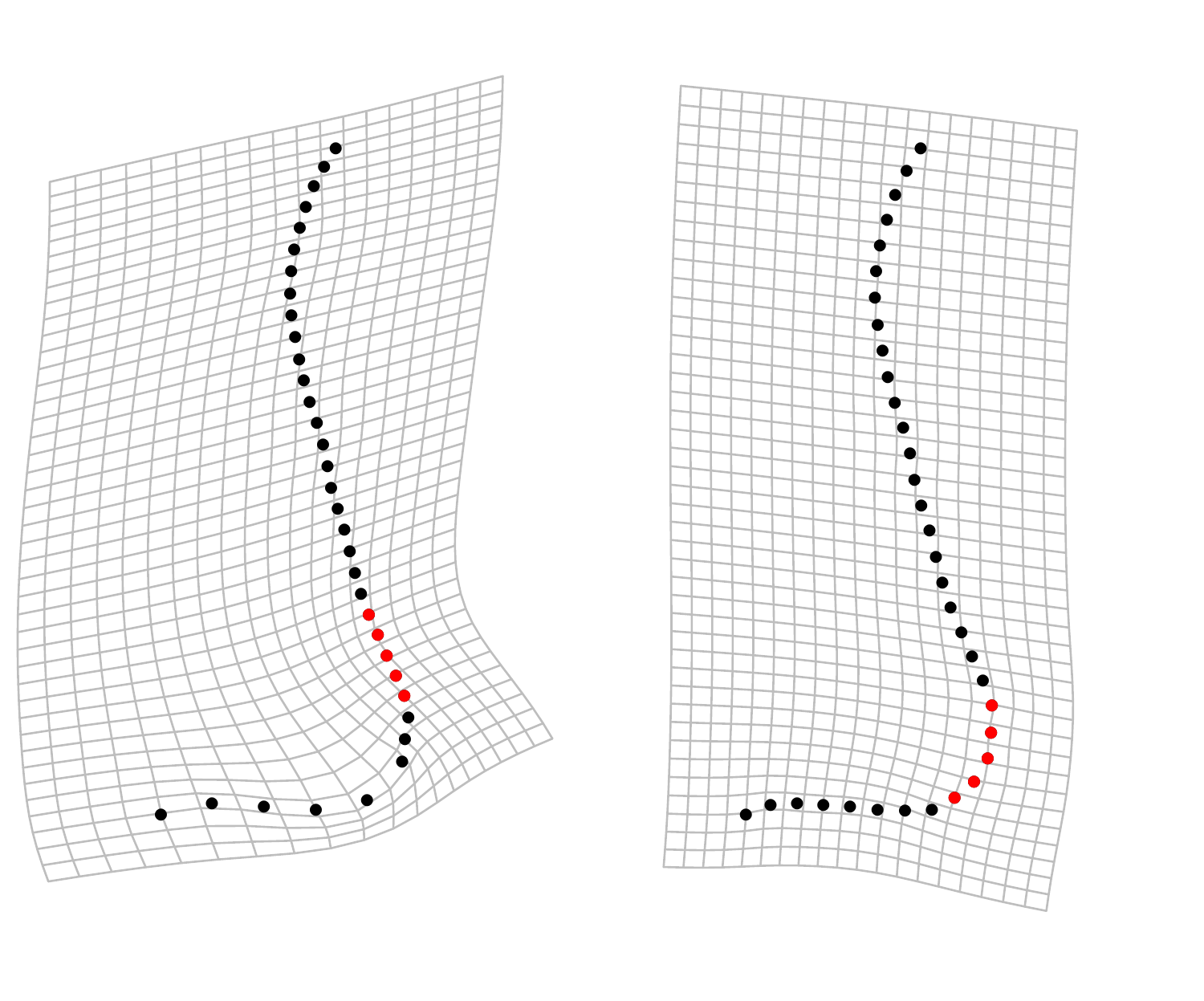}
}
\centerline{
   \includegraphics[width= 0.32\textwidth]{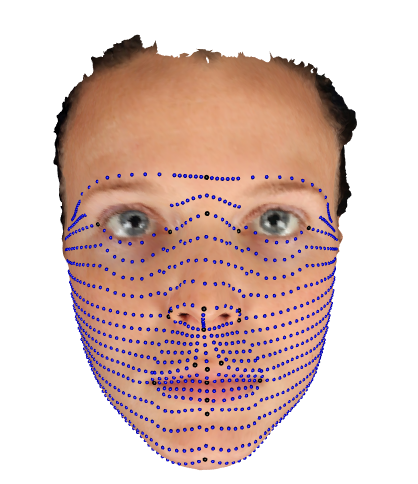}
   \hspace{3mm}
   \includegraphics[width= 0.32\textwidth]{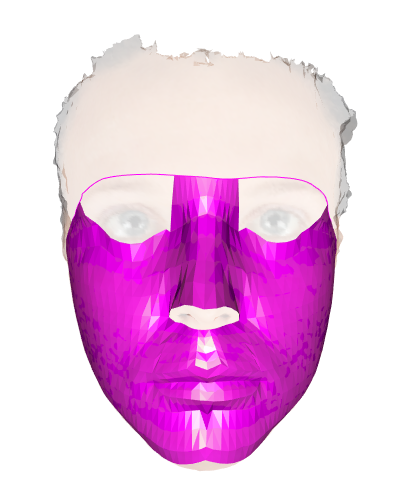}
}
\caption{The top panels show representations of the mid-line nasal profile curve, showing the template (left) and image before (centre) and after (right) sliding.  The red points show the nasal tip area on the template and the corresponding points on the image which sliding moves to the nasal tip area.  The lower images show the final facial model in $922$ discrete points (left) and fully rendered (right) forms.}
\label{figure:mask}
\end{figure}

An example of sliding across the nasal profile is shown in the top panels of Figure~\ref{figure:mask}.  For clarity of illustration, a two-dimensional curve projected onto the $x$-$z$ plane is shown, and an unequally-spaced discrete representation has been used.  The red points highlight those whose position in the ordered sequence on the template corresponds to the nasal tip region but which here lie much higher up on the nasal ridge.  The bending energy of the transformation between the image and the template is $7.37$.  The right hand panel shows the effects of iterative sliding, with the associated bending energy between the new image representation and the template considerably reduced to $0.68$.  Most importantly, the highlighted points now lie at the nasal tip and so a consistent interpretation of the curve shape has been constructed.


Any subsequent analysis will place central focus on the mean shape of the sample.  There is therefore particularly strong appeal in using the mean shape as the template against which the discrete representation of other images are adjusted.  This requires an iterative procedure where the Procrustes mean of the current set of images is used as the template for a second round of adjustment, the Procrustes mean is recomputed and the process repeated until convergence.  In practice, convergence is achieved very quickly after a few iterations.  In fact, the templates shown in Figure~\ref{figure:template} are mean shapes produced by this method.

The lower plots of Figure~\ref{figure:mask} show the end result of this process for an example image.  Notice that the eye area has been omitted due to the difficulties of locating curves in a region where image quality is often poor because of high reflectivity of the eyeball and the presence of eyelashes.  This problem could be addressed by warping curves onto the image from the template or by changing the data collection protocol to closed eyes.

\section{Evaluation of the performance of curve estimation}
\label{sec:evaluation}

The key step in the construction of the facial model described above is the estimation of the ridge and valley curves.  In order to evaluate the proposed method, three different approaches were employed.  The first uses Taylor's theorem and numerical methods to investigate the bias in estimating surface curvature, as this is the foundation of subsequent analysis.  The second method uses a simulation study to assess the performance of the estimated curves against known ridges.  In the third approach, estimated curves are compared with curves identified manually by a trained observer on a set of real images. 

\subsection{Bias in the estimation of surface curvature}
\label{subsection:nti}

As surface curvature is a function of eigenvalues and eigenvectors, there is little prospect of achieving an analytic description of the properties of estimators.  However, some theoretical investigation is feasible by numerical methods.  Consider a surface which has minimum (negative) curvature across a ridge and maximal (mildly positive or negative) curvature along the ridge.  The general parametric form for a surface, indexed by the parameters $(t_1, t_2) \in [0, 1]^2$, can be expressed in simple additive form as $(x(t_1), y(t_2), z_1(t_1) +  z_2(t_2))$, where $z_1(t_1)$ defines the cross-section (east-west) of the ridge and $z_2(t_2)$ defines the vertical movement (north-south) of the ridge curve.  The minimal curvature is generated by the curve $x(t_1), z_1(t_1))$ and has the form
$$
         \kappa_1(t_1) = \frac{x' z_1'' - z_1' x''}{(x'^2 + z_1'^2)^{\frac{3}{2}}} .
$$
\cite{koenderink} give the details.  The maximal curvature is generated by the curve $(y(t_2), z_2(t_2))$ in the same manner.  A further simplification reduces the surface to the form $(t_1, t_2, z_1(t_1))$, which specifies a curve across the ridge and imposes a constant path along it.  The minimal curvature across the ridge is then
$$
      \kappa_1(t_1) = \frac{z_1''}{ (1+ z_1'^2)^{\frac{3}{2}} },
$$
while the maximal curvature is zero everywhere.

The estimation of local surface curvature described in Section~\ref{sec:estimation} involves fitting a quadratic surface to a local neighbourhood.  The mean value of the parameter estimates from a linear model $Y=X \beta$ fitted to observed data $y$ with true underlying regression function $y = h(x)$ is $(X'X)^{-1}X'h(x)$.  The bias properties of curvature estimation are therefore provided by applying quadratic fitting to the true surface rather than the observed data.  Bias was therefore investigated by computing the true curvature numerically and comparing this with the curvature derived from fitting a local quadratic patch.  Different surface shapes, neighbourhood sizes and mesh resolutions were considered.

\begin{figure}[htp]
\centerline{
\includegraphics[width=0.5\textwidth]{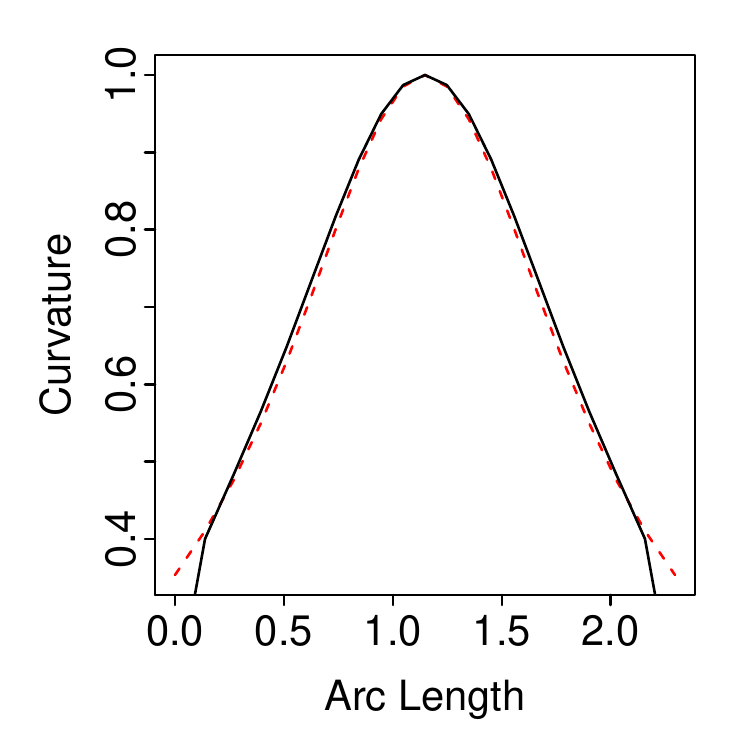}
\includegraphics[width=0.5\textwidth]{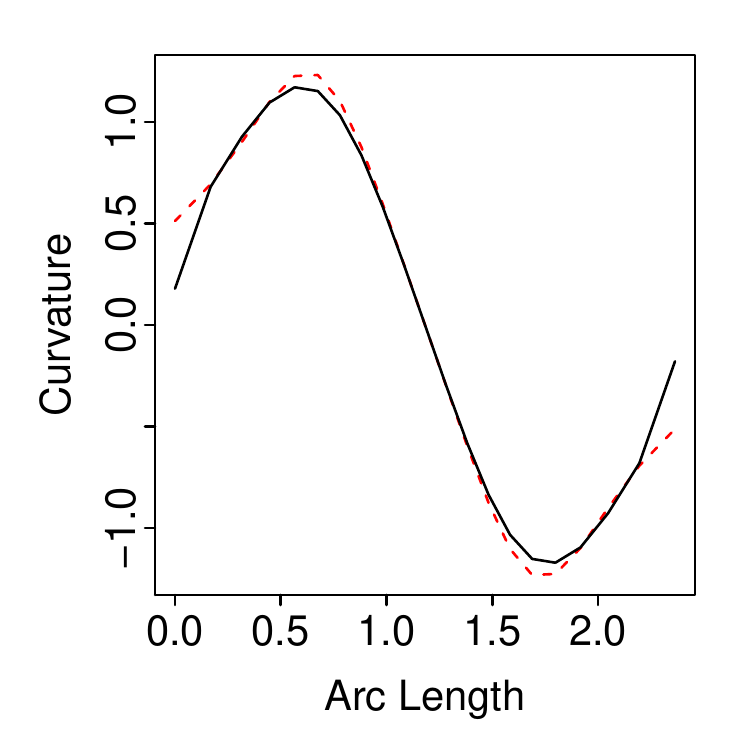}}
\caption{Plots of arc length and curvature for $z = -x^2$ (left) $z = -x^3$ (right) where the red dashed line represents the true curvature and the black full line represents the curvature computed from local quadratic patches.}
\label{figure:tayx3}
\end{figure}

Figure~\ref{figure:tayx3} shows examples of cross-sectional curvature based on the surfaces $z = -x^2$ (left) and $z = -x^3$ (right), where $x$ and $y$ lie on a regularly spaced grid of 1,681 points in $[0, 10]^2$ and a neighbourhood radius of $0.5$ was employed.  The differences between the true curvatures (red dashed lines) and those computed from quadratic patches (full black lines) are very minor, corresponding to small edge effects and to a slight underestimation of peak and trough curvature in the cubic example.  This example is typical of the cases considered, corresponding to surfaces and mesh resolutions similar to those encountered in facial surfaces.

\subsection{Simulation study}

\begin{figure}[htp]
\centerline{
\includegraphics[width=0.33\textwidth]{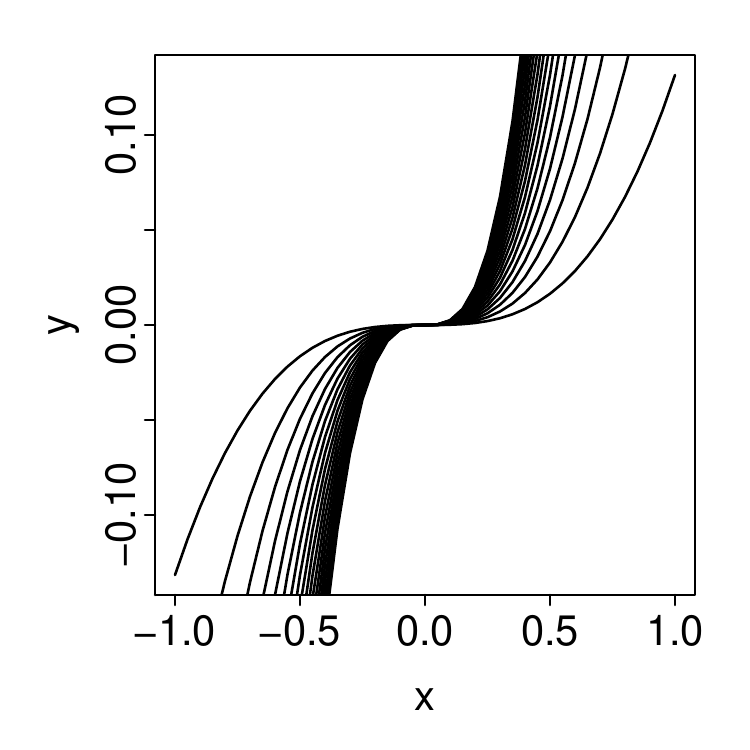}
\includegraphics[width=0.33\textwidth]{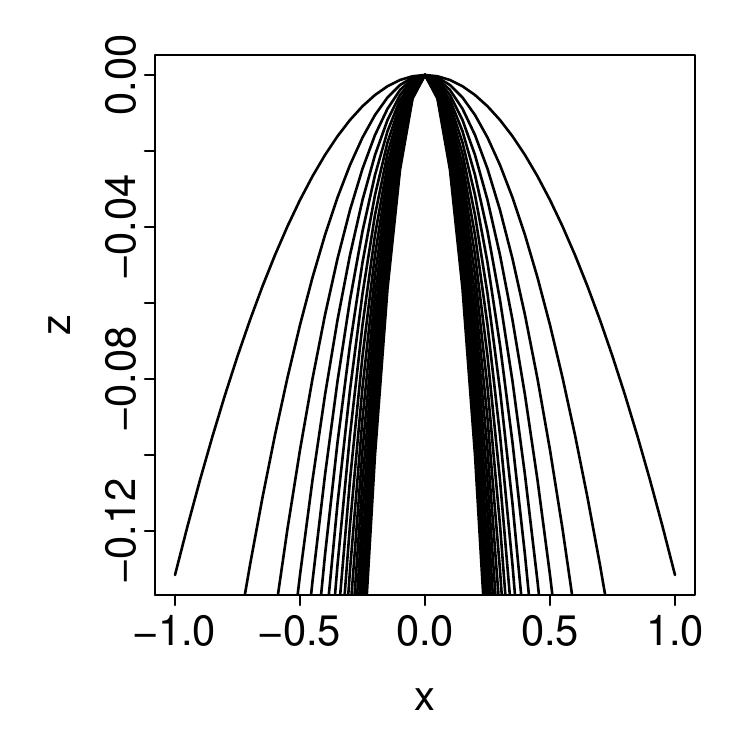}
\includegraphics[width=0.33\textwidth]{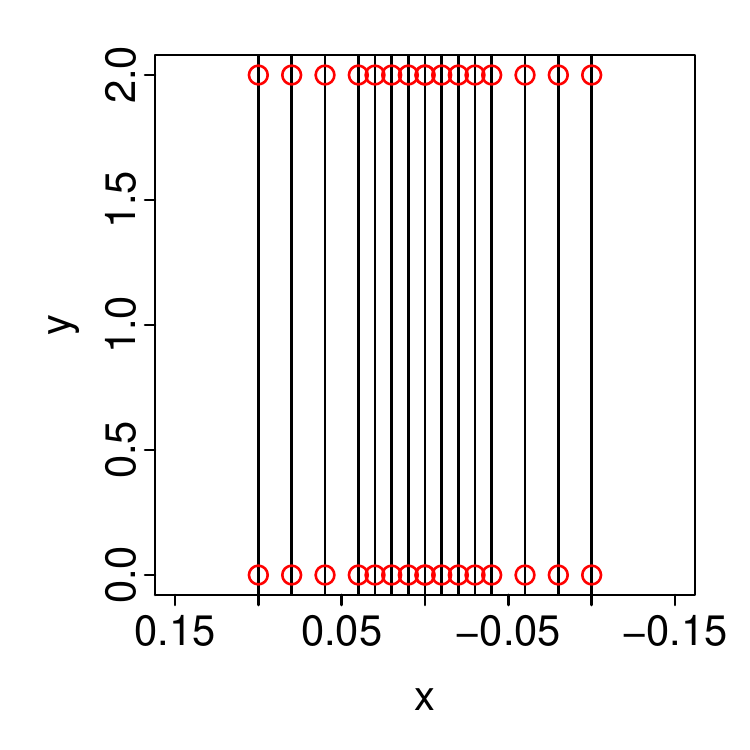}}
\caption{Ridge curve movement (left), cross-sectional ridge curvature (middle) and landmark inaccuracy from a $0$ reference (right).}
\label{figure:simul-image}
\end{figure}

Simulation studies were carried out to quantify the performance of ridge and valley curve estimation, using the surface construction described in subsection~\ref{subsection:nti}. This is based on the surface $z = -b\{x - c(y)\}^2$, where $x$ and $y$ lie on a regularly spaced grid of $41 \times 41 = 1,681$ points in $[0, 10]^2$ and $c(y)$ describes the location of the ridge curve in the $x-y$ plane.  The particular form $c(y) = (y/a)^\frac{1}{3}$ was used, so that the parameters $a$ and $b$ control the curvature of the ridge curve and the strength of the curvature across the ridge respectively.  Multiple neighbourhood radii were also employed.  The settings considered include:
\begin{center}
\begin{tabular}{rll}
\textit{ridge movement} & $a$ & curvature of the ridge curve, (0 -- 0.5)   \\
\textit{ridge curvature} &  $b$ & curvature of the ridge cross-section, (0 -- 0.5) \\
\textit{landmark inaccuracy} & $li$ & distance off the ridge curve, (0 -- 0.1) \\
\textit{mesh size} & $ms$ & distance between mesh points, (0.05 -- 0.1) \\
\textit{noise} & $\delta$ & uniform, added on the $z$-axis (0 -- 0.05)
\end{tabular}
\end{center}
The first three of these factors are displayed in Figure \ref{figure:simul-image}. In each case, $500$ simulations were generated and then each calculated curve represented by $21$ equally-spaced points along the ridge.  These were then compared with $21$ equally-spaced points along the true ridge curve and the separation quantified by average Euclidean distance.
The full results are available in the \textit{Supplementary material}.  In summary, the simulations show very good performance in estimation with, as expected, a slight deterioration with increases in each of the factors listed above.  

\subsection{Manual Comparison}

Manual identification by trained individuals has been regarded as the `gold standard' for the production of anatomical landmarks.  It needs to be recognised that manual methods are also subject to variation and inaccuracy, and that this will undoubtedly increase in the manual identification of curves rather than points.  Nonetheless, it is valuable to compare the curves produced by the facial model described above with those identified manually, here using the  \textit{Landmark} \copyright{IDAV} software.  The facial model curves and the manually marked curves were constructed on $55$ images and represented by sets of discrete points to allow Procrustes matching of each pair of curves.  This gave an overall root mean square distance between points of $1.70\,$mm.  Differences between the curves were strongest in the lip borders and the nasal bridge.  When these regions are omitted, the overall root mean square distance between curves drops to $0.60\,$mm. 

It is instructive to consider the reasons why these two areas show the largest differences.  When marking lips, manual observers are likely to be strongly influenced by colour change at the so-called `vermillion border'.  However, it is not always the case that the ridge of the lips coincides exactly with this colour change.  At the nasal bridge, there is a tendency for manual observers to mark positions which are slightly `north' of the ridge curve.  It can be very difficult to make a visual judgement on where this ridge lies, with a great deal of examination of the image required at different orientations.  In contrast, the facial model employs direct information on surface curvature, unaffected by perceptual considerations.



\section{An application to sexual dimorphism}
\label{sec:application}

The differences in facial shape between males and females have been studied from a variety of perspectives.  For example, research in psychology \citep{bruce1993sex, Armann201269} has used reported male/female visual judgement to identify both the key facial features which differentiate the sexes as well as those features used to make gender classifications.  Shape information has traditionally been quantified through distances, angles and ratios computed from landmarks in two- and three-dimensional images.  Classification algorithms are estimated to be $94$\% accurate when combining two- and three-dimensional information.  However, the faces misclassified by the algorithms are not the same ones as those misclassified by the human observers.  \cite{kramer2012lack} used two-dimensional photographs and three-dimensional scans to measure facial width-to-height ratio for men and women, using landmarks in the middle of the face, but found no evidence of sexual dimorphism.

The manner in which shape is quantified clearly has a very big influence on the information which is then available for analysis.  In particular, the shape information present in landmarks is necessarily limited.  Recent examples of full three-dimensional surface representations of faces include \cite{srivastava2011shape}, who use elastic deformations and level curves, and \cite{claes2012sexual} who map a `facial mask' onto the images and focus particularly on symmetric/asymmetric components of sexual dimorphism.  The facial model developed in the present paper was used to tackle the general question of sexual dimorphism, with particular interest in the information exploited by the different levels of representations in landmarks, curves and surfaces.

A sample of $250$ facial images was used.  All statistical tests reported in this section compare males and females through a permutation approach, with $1000$ permutations of sex labels, to avoid reliance on distributional assumptions for very high-dimensional data.  A larger mean size should be expected for males and this was confirmed by a two sample $t$-test ($p = 0.011$).  Further analysis was based on principal components, which provide a standard mechanism to describe the main modes of shape variation in a smaller number of dimensions, as discussed by \cite{dryden-1998-book} for example.  The first $10$ principal components were used here are this captures around $90$\% of the shape variation.  A Hotelling's $T^2$ test of differences in mean facial shape between males and females, based on these component, found a marginally significant difference of $p = 0.048$.  However, this was based on the full facial mesh and a more informative analysis is likely to be achieved by examination of the individual components.

\begin{figure}
  \centerline{
   \includegraphics[width= 0.5\textwidth]{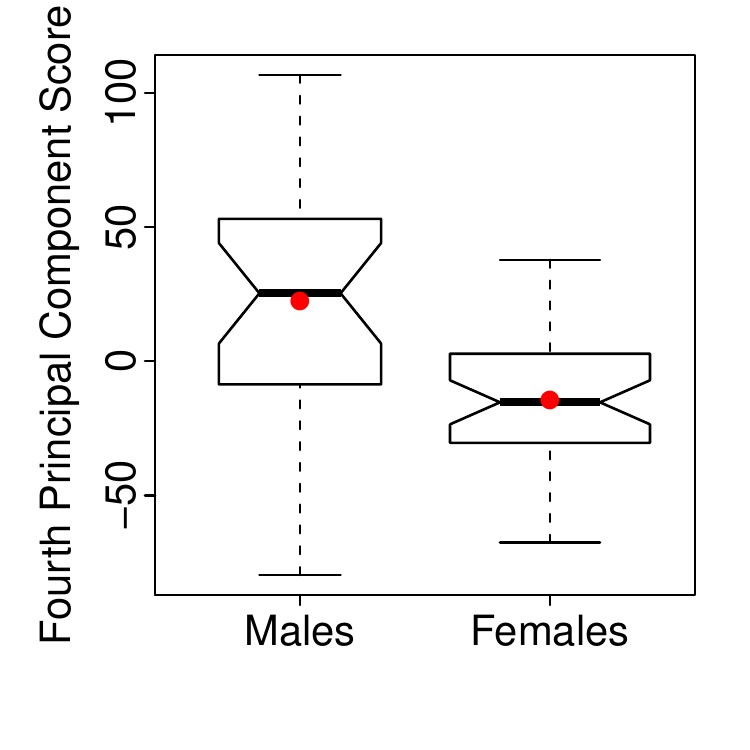}
   \includegraphics[width= 0.5\textwidth]{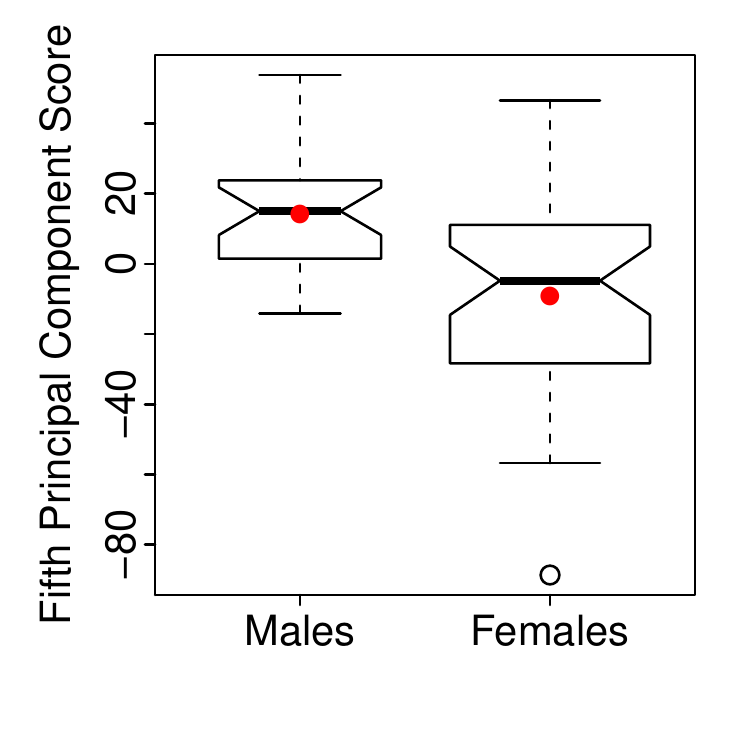}}
\centerline{
   \includegraphics[width= 0.5\textwidth]{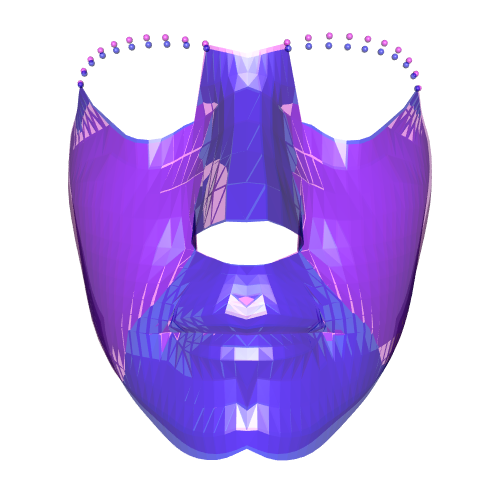}
   \includegraphics[width= 0.5\textwidth]{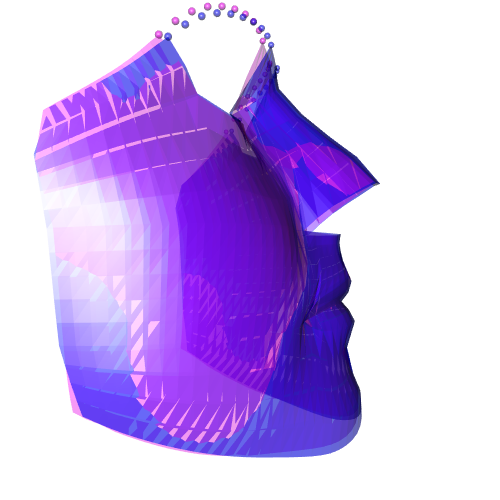}}
   \caption{The upper panels show boxplots and means (red point) of the 4th (brow) and 5th (cheek) principal components, separated by males and females.  The lower images illustrate sexual dimorphism through the mean facial shapes of males (blue) and females (pink).}
\label{figure:sexual-dimorphism}
\end{figure}

The first few principal components capture large-scale variation, for example in height-width ratio, which is common to males and females.  However, the upper plots of Figure~\ref{figure:sexual-dimorphism} show clear evidence of sexual dimorphism in the 4th and 5th principal components derived from the facial surface models (p-values $0.009$ and $0.008$ respectively).  When these components are re-expressed as shape changes, they correspond to protrusion of the brow ridge (stronger in males) and protrusion of the cheekbones (stronger in females).  The mean surface shape models for males and females are displayed in Figure~\ref{figure:sexual-dimorphism}, using transparent surface representations.  The cheek protrusion in females is particularly noticeable with a compensating protrusion in the nasal and mandible regions in males.  Differences in the brow ridge are also evident, with males exhibiting a lower ridge with greater protrusion.

If analysis is based on the facial curves then again the fourth principal component is associated with the protrusion of the brow ridge, and this is significantly different for males and females.  There is no natural ridge or valley curve across the cheek so this manifestation of sexual dimorphism cannot be captured.   However, if landmarks alone are used for analysis then no significant differences are identified.  This highlights the additional information expressed in the full facial surface model in contrast with the traditional landmark approach.  It also signals the key role played by the facial curves, where much of the information in facial shape resides.

\section{Discussion}
\label{sec:discussion}

The model described in the paper gives a full description of a three-dimensional manifold through estimation of its ridge and valley curves and subsequent `in-filling' of the intervening surface patches.  The methods used respect the inherently two-dimensional nature of the manifold by operating in relevant two-dimensional spaces.  The maximisation of curvature along path integrals is used as the guiding principle for the estimation of the ridge and valley curves, with efficient implementation through p-spline representations.  The application to facial surfaces illustrates the additional information present in a surface, rather than landmark, representation.  The fact that a substantial component of this additional information is captured by the facial curves highlights the central role played by this key shape structure.

The analysis here has exploited the presence of traditional landmarks by using these to `seed' the construction of the ridge and valley curves.  The existence of methods of estimating these curves opens up the possibility of informing on the location of landmarks, which can be defined in terms of curve crossings or points of extreme geodesic curvature.  The combined estimation of curves and landmarks is therefore a natural topic for further research.

\section*{Acknowledgement}

The work of Adrian Bowman and Stanislav Katina was supported by a Wellcome Trust grant (WT086901MA) to the Face3D research consortium (\texttt{www.Face3D.ac.uk}), under whose auspices the facial data were collected.

\bibliographystyle{Chicago}
\bibliography{arxiv.bib}

\begin{thebibliography}{}

\bibitem[\protect\citeauthoryear{Armann and Balthoff}{Armann and
  Balthoff}{2012}]{Armann201269}
Armann, R. and Balthoff, I. (2012).
\newblock Male and female faces are only perceived categorically when linked to
  familiar identities. and when in doubt, he is a male.
\newblock {\em Vision Research\/}~{\em 63\/}(0), 69--80.

\bibitem[\protect\citeauthoryear{Bookstein}{Bookstein}{1997}]{bookstein-1997-mia}
Bookstein, F.~L. (1997).
\newblock Landmark methods for forms without landmarks: morphometrics of group
  differences in outline shape.
\newblock {\em Medical image analysis\/}~{\em 1\/}(3), 225--243.

\bibitem[\protect\citeauthoryear{Bookstein et~al.}{Bookstein
  et~al.}{1989}]{bookstein-1989-ieeepami}
Bookstein, F.~L. et~al. (1989).
\newblock Principal warps: Thin-plate splines and the decomposition of
  deformations.
\newblock {\em IEEE Transactions on pattern analysis and machine
  intelligence\/}~{\em 11\/}(6), 567--585.

\bibitem[\protect\citeauthoryear{Bowman, Katina, Smith, and Brown}{Bowman
  \textit{et~al.}}{2015}]{bowman-2015-csda}
Bowman, A.~W., Katina, S., Smith, J., and Brown, D. (2015).
\newblock Anatomical curve identification.
\newblock {\em Computational Statistics \& Data Analysis\/}~{\em 86}, 52--64.

\bibitem[\protect\citeauthoryear{Bruce, Burton, Hanna, Healey, Mason, Coombes,
  Fright, and Linney}{Bruce \textit{et~al.}}{1993}]{bruce1993sex}
Bruce, V., Burton, A.~M., Hanna, E., Healey, P., Mason, O., Coombes, A.,
  Fright, R., and Linney, A. (1993).
\newblock Sex discrimination: how do we tell the difference between male and
  female faces?
\newblock {\em Perception\/}.

\bibitem[\protect\citeauthoryear{Che, Zhang, Zhang, Paul, and Xu}{Che
  \textit{et~al.}}{2011}]{che2011ridge}
Che, W., Zhang, X., Zhang, Y., Paul, J., and Xu, B. (2011).
\newblock Ridge extraction of a smooth 2-manifold surface based on vector
  field.
\newblock {\em Computer Aided Geometric Design\/}~{\em 28\/}(4), 215--232.

\bibitem[\protect\citeauthoryear{Claes, Walters, Shriver, Puts, Gibson,
  Clement, Baynam, Verbeke, Vandermeulen, and Suetens}{Claes
  \textit{et~al.}}{2012}]{claes2012sexual}
Claes, P., Walters, M., Shriver, M.~D., Puts, D., Gibson, G., Clement, J.,
  Baynam, G., Verbeke, G., Vandermeulen, D., and Suetens, P. (2012).
\newblock Sexual dimorphism in multiple aspects of 3d facial symmetry and
  asymmetry defined by spatially dense geometric morphometrics.
\newblock {\em Journal of anatomy\/}~{\em 221\/}(2), 97--114.

\bibitem[\protect\citeauthoryear{De~Berg, Van~Kreveld, Overmars, and
  Schwarzkopf}{De~Berg \textit{et~al.}}{2000}]{deberg-2000-book}
De~Berg, M., Van~Kreveld, M., Overmars, M., and Schwarzkopf, O.~C. (2000).
\newblock {\em Computational geometry}.
\newblock New York: Springer.

\bibitem[\protect\citeauthoryear{Dryden and Mardia}{Dryden and
  Mardia}{1998}]{dryden-1998-book}
Dryden, I.~L. and Mardia, K. (1998).
\newblock {\em Statistical shape analysis}.
\newblock New York: Wiley.

\bibitem[\protect\citeauthoryear{Eilers and Marx}{Eilers and
  Marx}{1996}]{eilers1996flexible}
Eilers, P.~H. and Marx, B.~D. (1996).
\newblock Flexible smoothing with b-splines and penalties.
\newblock {\em Statistical science\/}, 89--102.

\bibitem[\protect\citeauthoryear{Farkas}{Farkas}{1994}]{farkas-1994-book}
Farkas, L. (1994).
\newblock {\em Anthropometry of the head and face\/} (Second ed.).
\newblock New York: Raven Press.

\bibitem[\protect\citeauthoryear{Goldfeather and Interrante}{Goldfeather and
  Interrante}{2004}]{goldfeather2004novel}
Goldfeather, J. and Interrante, V. (2004).
\newblock A novel cubic-order algorithm for approximating principal direction
  vectors.
\newblock {\em ACM Transactions on Graphics (TOG)\/}~{\em 23\/}(1), 45--63.

\bibitem[\protect\citeauthoryear{Gunz, Mitteroecker, and Bookstein}{Gunz
  \textit{et~al.}}{2005}]{gunz-2005-inbook}
Gunz, P., Mitteroecker, P., and Bookstein, F.~L. (2005).
\newblock Semilandmarks in three dimensions.
\newblock In {\em Modern morphometrics in physical anthropology}, pp.\  73--98.
  Springer.

\bibitem[\protect\citeauthoryear{Hammond, Hutton, Allanson, Campbell, Hennekam,
  Holden, Patton, Shaw, Temple, Trotter, Murphy, and Winter}{Hammond
  \textit{et~al.}}{2004}]{hammond20043d}
Hammond, P., Hutton, T.~J., Allanson, J.~E., Campbell, L.~E., Hennekam, R.,
  Holden, S., Patton, M.~A., Shaw, A., Temple, I.~K., Trotter, M., Murphy,
  K.~C., and Winter, R.~M. (2004).
\newblock 3d analysis of facial morphology.
\newblock {\em American journal of medical genetics Part A\/}~{\em 126\/}(4),
  339--348.

\bibitem[\protect\citeauthoryear{Hastie and Stuetzle}{Hastie and
  Stuetzle}{1989}]{hastie-1989-jasa}
Hastie, T. and Stuetzle, W. (1989).
\newblock Principal curves.
\newblock {\em Journal of the American Statistical Association\/}~{\em
  84\/}(406), 502--516.

\bibitem[\protect\citeauthoryear{Katina, McNeil, Ayoub, Guilfoyle, Khambay,
  Siebert, Sukno, Rojas, Vittert, Waddington, Whelan, and Bowman}{Katina
  \textit{et~al.}}{2016}]{katina-2015-janatomy}
Katina, S., McNeil, K., Ayoub, A., Guilfoyle, B., Khambay, B., Siebert, P.,
  Sukno, F., Rojas, M., Vittert, L., Waddington, J., Whelan, P.~F., and Bowman,
  A.~W. (2016).
\newblock The definitions of three-dimensional landmarks on the human face: an
  interdisciplinary view.
\newblock {\em Journal of Anatomy\/}~{\em 228\/}(3), 355--365.

\bibitem[\protect\citeauthoryear{Kent, Mardia, and West}{Kent
  \textit{et~al.}}{1996}]{kent-1996-bmvc}
Kent, J.~T., Mardia, K.~V., and West, J. (1996).
\newblock Ridge curves and shape analysis.
\newblock In {\em BMVC}, pp.\  1--10.

\bibitem[\protect\citeauthoryear{Kneip and Ramsay}{Kneip and
  Ramsay}{2008}]{kneip-2008-jasa}
Kneip, A. and Ramsay, J.~O. (2008).
\newblock Combining registration and fitting for functional models.
\newblock {\em Journal of the American Statistical Association\/}~{\em
  103\/}(483), 1155--1165.

\bibitem[\protect\citeauthoryear{Koenderink}{Koenderink}{1990}]{koenderink-1990-book}
Koenderink, J. (1990).
\newblock {\em Solid shape}, Volume~2.
\newblock Cambridge: Cambridge Univ Press.

\bibitem[\protect\citeauthoryear{Koenderink and Doorn}{Koenderink and
  Doorn}{1992}]{koenderink}
Koenderink, J. and Doorn, A. ({1992}).
\newblock {Surface shape and curvature scales}.
\newblock {\em {IVC}\/}~{\em {10}}, {557--565}.

\bibitem[\protect\citeauthoryear{Koenderink and van Doorn}{Koenderink and van
  Doorn}{1992}]{koenderink-1992-imagevisioncomputing}
Koenderink, J. and van Doorn, A. (1992).
\newblock Surface shape and curvature scales.
\newblock {\em Image and vision computing\/}~{\em 10\/}(8), 557--564.

\bibitem[\protect\citeauthoryear{Kramer, Jones, and Ward}{Kramer
  \textit{et~al.}}{2012}]{kramer2012lack}
Kramer, R.~S., Jones, A.~L., and Ward, R. (2012).
\newblock A lack of sexual dimorphism in width-to-height ratio in white
  european faces using 2d photographs, 3d scans, and anthropometry.
\newblock {\em PloS one\/}~{\em 7\/}(8), e42705.

\bibitem[\protect\citeauthoryear{Meyer, Barr, Lee, and Desbrun}{Meyer
  \textit{et~al.}}{2002}]{meyer2002generalized}
Meyer, M., Barr, A., Lee, H., and Desbrun, M. (2002).
\newblock Generalized barycentric coordinates on irregular polygons.
\newblock {\em Journal of graphics tools\/}~{\em 7\/}(1), 13--22.

\bibitem[\protect\citeauthoryear{Ohtake, Belyaev, and Seidel}{Ohtake
  \textit{et~al.}}{2004}]{ohtake2004ridge}
Ohtake, Y., Belyaev, A., and Seidel, H.-P. (2004).
\newblock Ridge-valley lines on meshes via implicit surface fitting.
\newblock In {\em ACM Transactions on Graphics (TOG)}, Volume~23, pp.\
  609--612. ACM.

\bibitem[\protect\citeauthoryear{Patrangenaru and Ellingson}{Patrangenaru and
  Ellingson}{2015}]{patrangenaru-2015-book}
Patrangenaru, V. and Ellingson, L. (2015).
\newblock Nonparametric statistics on manifolds and their applications to
  object data analysis.

\bibitem[\protect\citeauthoryear{Rohr}{Rohr}{2001}]{rohr-2001-book}
Rohr, K. (2001).
\newblock {\em Landmark-based image analysis: using geometric and intensity
  models}, Volume~21.
\newblock Springer Science \& Business Media.

\bibitem[\protect\citeauthoryear{Srivastava, Klassen, Joshi, and
  Jermyn}{Srivastava \textit{et~al.}}{2011}]{srivastava2011shape}
Srivastava, A., Klassen, E., Joshi, S.~H., and Jermyn, I.~H. (2011).
\newblock Shape analysis of elastic curves in euclidean spaces.
\newblock {\em Pattern Analysis and Machine Intelligence, IEEE Transactions
  on\/}~{\em 33\/}(7), 1415--1428.

\bibitem[\protect\citeauthoryear{Srivastava, Samir, Joshi, and
  Daoudi}{Srivastava \textit{et~al.}}{2009}]{srivastava2009elastic}
Srivastava, A., Samir, C., Joshi, S.~H., and Daoudi, M. (2009).
\newblock Elastic shape models for face analysis using curvilinear coordinates.
\newblock {\em Journal of Mathematical Imaging and Vision\/}~{\em 33\/}(2),
  253--265.

\bibitem[\protect\citeauthoryear{Stylianou and Farin}{Stylianou and
  Farin}{2004}]{stylianou-2004-ieee}
Stylianou, G. and Farin, G. (2004).
\newblock Crest lines for surface segmentation and flattening.
\newblock {\em IEEE Transactions on Visualization \& Computer Graphics\/}~(5),
  536--544.

\bibitem[\protect\citeauthoryear{Sukno, Waddington, and Whelan}{Sukno
  \textit{et~al.}}{2014}]{sukno-2014-cybernetics}
Sukno, F.~M., Waddington, J.~L., and Whelan, P.~F. (2014).
\newblock 3d facial landmark localization with asymmetry patterns and shape
  regression from incomplete local features.
\newblock {\em IEEE Transactions on Cybernetics\/}.

\bibitem[\protect\citeauthoryear{Zhao, Dellandrea, Chen, and Kakadiaris}{Zhao
  \textit{et~al.}}{2011}]{zhao-2011-smc}
Zhao, X., Dellandrea, E., Chen, L., and Kakadiaris, I. (2011).
\newblock Accurate landmarking of three-dimensional facial data in the presence
  of facial expressions and occlusions using a three-dimensional statistical
  facial feature model.
\newblock {\em Systems, Man, and Cybernetics, Part B: Cybernetics, IEEE
  Transactions on\/}~{\em 41\/}(5), 1417--1428.

\end{thebibliography}

\end{document}